\documentclass[final,iicol,natbib]{sn-jnl}
\usepackage[round]{natbib}
\usepackage{breakcites}

\jyear{2021}%

\theoremstyle{thmstyleone}%
\theoremstyle{thmstyletwo}%

\theoremstyle{thmstylethree}%

\begin{document}

\title[Article Title]{IoT Device Identification Based on Network Communication Analysis Using Deep Learning}


\author*[]{\fnm{Jaidip} \sur{Kotak*}}\email{jaidip@post.bgu.ac.il}

\author[]{\fnm{Yuval} \sur{Elovici}}\email{elovici@bgu.ac.il}

\affil[]{\orgdiv{Department of Software and Information Systems Engineering}, \orgname{Ben-Gurion University of the Negev}, \orgaddress{\city{Beer-Sheva}, \country{Israel}}}


\abstract{Attack vectors for adversaries have increased in organizations because of the growing use of less secure IoT devices. The risk of attacks on an organization's network has also increased due to the bring your own device (BYOD) policy which permits employees to bring IoT devices onto the premises and attach them to the organization's network. To tackle this threat and protect their networks, organizations generally implement security policies in which only white-listed IoT devices are allowed on the organization's network. To monitor compliance with such policies, it has become essential to distinguish IoT devices permitted within an organization's network from non-white-listed (unknown) IoT devices.  In this research, deep learning is applied to network communication for the automated identification of IoT devices permitted on the network. In contrast to existing methods, the proposed approach does not require complex feature engineering of the network communication, because the 'communication behavior' of IoT devices is represented as small images which are generated from the device’s network communication payload.  The proposed approach is applicable for any IoT device, regardless of the protocol used for communication. As our approach relies on the network communication payload, it is also applicable for the IoT devices behind a network address translation (NAT) enabled router. In this study, we trained various classifiers on a publicly accessible dataset to identify IoT devices in different scenarios, including the identification of known and unknown IoT devices, achieving over 99\% overall average detection accuracy.}




\maketitle

\section{Introduction}\label{sec1}

\footnotetext{This is an extended and revised version of the paper \citep{kotak2019iot} that was presented at the CISIS 2020 conference and was published in its proceedings.}
The term “Internet of Things” is described as a collection of devices having low computing capabilities and sensing and/or actuating abilities that extend Internet connectivity beyond that of normal devices like smartphones, laptops, and computers. IoT devices are widely used in different domains like medical, manufacturing, environmental sustainability, security, etc. \citep{sangaiah2020cognitive,sangaiah2020iot,sangaiah2019enforcing}.  The total number of IoT devices has already exceeded the number of the humans on the planet, and by 2025, the amount of IoT devices is anticipated to reach around 75.44 billion worldwide \citep{vailshery_2016}. 

\textbf{Cause of the problem:} The absence of governmental standards for IoT devices and the limited security awareness of device manufacturers, vendors, and users have left IoT devices vulnerable \citep{interpol}. When these IoT devices are attached to organizational networks, they increase the exposure of these networks to adversaries. With the help of search engines like Shodan \citep{shodan}, adversaries can locate IoT devices with specific configurations and target them due to their less secure nature. In a targeted attack on the network, adversaries can leverage IoT devices that are part of the organizational network to perform the specific task of the ongoing attack \citep{arenson_2018,anthraper2019security}. Therefore, it has become essential to identify the IoT devices permitted on the organizational network to reduce the threat and increase the security of the network.

\textbf{Problem statement:} It is difficult for organizations to identify the IoT devices on their networks. The threat to the organizational network increases in organizations that have adopted the bring your own device (BYOD) policy whereby employees’ IoT devices are not only allowed on the premises but can be connected to the organizational network \citep{olalere2015review,abomhara2015cyber,andrea2015internet}. It is also difficult to perform investigations after an attack using traditional forensic methodologies, because different IoT devices use different protocols to communicate \citep{kotak2019comparative,shah2019memory}. To address these challenges, organizations need a means of identifying the IoT devices (both known and unknown to network administrators) connected to their networks. This can help organizations effectively manage the security issues associated with IoT devices and determine whether the behavior/activity of the connected IoT devices is normal.

\textbf{Existing approaches \& their limitations:} Prior research suggested ways of identifying IoT devices by analyzing the network communication \citep{meidan2017profiliot,meidan2017detection,sivanathan2017characterizing,aksoy2019automated,miettinen2017iot,sivanathan2018classifying,yu2020you,meidan2019privacy}. Since the methods proposed are based on machine learning, feature engineering (i.e., feature extraction, selection, and tuning) is required. This necessitates manual input from subject matter experts, which is both expensive and prone to errors. Existing approaches are time-consuming, requiring multiple sessions to identify known and unknown (also referred to as unauthorized in this paper) IoT devices, and tend to have complex architecture, since they use multistage models. In addition, most existing approaches are not applicable when the IoT devices are behind a NAT (network address translation) enabled router, as many of the features get altered in the NAT process \citep{meidan2019privacy}. 

\textbf{Proposed solution:} Our approach mitigates these limitations. Just a single session is needed to identify known and unknown IoT devices; in addition, it is free from the burden of feature engineering and the errors that are accompanied with feature engineering, and it has a simple architecture. The proposed approach also applies to IoT devices behind a NAT-enabled router, as the NAT process does not alter the payload of the communication.

The proposed approach enables us to identify known and unknown IoT devices in the network in various scenarios.  Organizations’ use of DHCP and the ease in which MAC addresses can be spoofed has made it difficult to identify IoT devices using traditional approaches \citep{xiao2018iot,ling2017security}. Therefore, rather than focusing on the header of the packets, our approach focuses on the TCP content of the packets \citep{meidan2017profiliot,meidan2017detection,sivanathan2018classifying,yu2020you}. Compared to previously proposed approaches for identifying IoT devices in organizational networks, our approach is a less complex and more generic solution with equal or greater accuracy. This paper is an extended version of a paper \citep{kotak2019iot} published at the CISIS 2020 conference. This extended version contains three additional experiments demonstrating the wide applicability of the proposed approach in various scenarios and discusses about its feasibility when IoT devices are behind NAT-enabled router. This version of the paper also contains the detailed results of the experiments performed.

\textbf{Contribution: }The contributions of our research are as follows:
\begin{itemize}

\item{To the best of our knowledge, we are the first to classify and identify IoT devices by applying deep learning techniques on the TCP payload of network communication.}
\item{The proposed approach can be used to differentiate IoT and non-IoT devices.}
\item{The proposed approach can be used to identify the communication of a particular IoT device in the network communication.}
\item{The proposed approach can be used to identify white-listed IoT devices in the network communication.}
\item{The proposed approach is also applicable for identifying IoT devices behind a NAT.}
\item{Unlike exiting approaches which need multiple TCP sessions to detect an IoT device, our approach needs just a single TCP session to detect the source IoT device.}
\item{The proposed approach has no feature engineering burden and a simple architecture.}

\end{itemize}

\section{Related Work}\label{sec2}

Researchers have used machine learning to classify network communication for the purpose of identifying the services being used on the source computers \citep{wang2015applications,lopez2017network}; transfer learning techniques, in particular, have been used for network communication classification with promising results \citep{sun2018network}. Deep learning and machine learning algorithms have been used to differentiate benign and malicious communication \citep{wang2017malware,celik2015malware}. Another study \citep{acar2020peek} showed how adversaries can use machine learning techniques to automatically identify user activities from the network communication of residential IoT devices. Following the brief overview of how machine learning has been used for network communication classification provided above, we now present a detailed discussion on the use of various machine learning-based approaches for IoT device identification.

\citep{meidan2017profiliot} applied machine learning on the features of a TCP session to identify different IoT device classes. Features that were considered as an input to machine learning models were from the network, transport, and application layers of the TCP session, along with additional information from Alexa Rank \citep{alexa} and GeoIP \citep{geoip.com} for IP addresses. As part of experiments, the researchers classified nine different IoT devices by utilizing various machine learning classifiers. An optimal threshold value was obtained for each classifier to identify the source IoT device for the given instance. The threshold value for each classifier was the total number of sequences of TCP sessions required to identify the source IoT device. Despite over 99\% accuracy achieved in identifying IoT devices, their approach was only effective on IoT devices using HTTP and TLS protocols, as features from the application layer were considered for just these two protocols. Their proposed solution also had limitations in that it requires different types of machine learning models and different threshold values (numbers of TCP session sequences) to identify different IoT devices.

In \citep{meidan2017detection}, the authors utilized machine learning techniques to identify white-listed IoT devices in the network communication. A total of 10 IoT devices were used, and in each experiment, one IoT device from the white-list was removed and considered an unknown IoT device. Statistical features and features from different network layers were used to train the model.  The 10 most influential features were identified from the over 300 features used, and most of them were found to be statistical features that were obtained from the TTL (time-to-live) value. The average accuracy was 99\% for the correct classification of white-listed IoT devices. Because the work was an extended version of \citep{meidan2017profiliot}, it inherits its limitations (as explained above).

In \citep{sivanathan2017characterizing}, the authors proposed the use of statistical attributes of network traffic such as data rates and burstiness, activity cycles, signaling patterns, and other features, to characterize IoT devices, including their typical behavior mode. Twenty IoT devices, along with non-IoT device traffic, were used in the experiment. The study’s authors developed a classification method that can distinguish IoT from non-IoT traffic and identify specific IoT devices with over 95\% accuracy. The limitation of the proposed approach is that it requires domain experts to identify the features.

The authors of \citep{aksoy2019automated} introduced a system for the automated classification of device characteristics, System IDentifier (SysID), based on the network traffic of IoT devices. SysID identifies network traffic based on any single packet originating from the IoT device. The authors used a genetic algorithm (GA) to determine relevant features in different protocol headers and then used various ML algorithms for identification of IoT devices. The protocol headers of the TCP, UDP, and DNS protocols were used to extract features. The traffic of 23 IoT devices traffic was used. Their evaluation showed that SysID was able to identify the IoT device type from a single packet with over 95\% accuracy. The limitation of the work stems from the fact that header fields like port numbers from TCP and UDP protocol headers were used – such features are less reliable, since vendors can change the port number for communication.  

In \citep{miettinen2017iot}, the authors proposed IoT SENTINEL an automated device-type identification method for security enforcement in the IoT. They used 23 features extracted from 16 different protocols and concatenated these features for 12 packets, resulting in a single input instance of 276 dimensions (12 packets × 23 features). In the experiment 27 IoT devices were examined, however the average accuracy obtained was 81.5\%, as the proposed approach was not effective in identifying different devices from the same vendors.  

\citep{sivanathan2018classifying} proposed a machine learning model based on the statistical attributes of network communication, such as cipher suites, signaling patterns, activity cycles, and port numbers. A multistage model was proposed in their research, which also considers flow level features, like the NTP interval, DNS interval, sleep time, flow rate, flow duration, and flow volume extracted from the network communication, to classify the IoT devices. The authors used 28 IoT devices and one non-IoT device class label in their experiment and achieved over 99\% accuracy. However, the proposed approach has limitations in that domain experts are required in order to identify which features to use; in addition, the proposed method has a complicated multistage architecture. In addition, its reliance on features like domains and port numbers is inadvisable, since they can be changed by vendors at any time.

\citep{yu2020you} proposed a multi-view wide and deep learning model to identify IoT devices. Their approach is limited to the identification of IoT devices connected to a Wi-Fi network.  Network communication features from passively received broadcast and multicast (BC/MC) packets were extracted for use in training the model. The authors extracted three categories of features from the sniffed BC/MC packets: (1) the identifiers that are unique to each IoT device; (2) the main features needed for robust discriminators (i.e., those features that can be combined to collectively provide ample information to differentiate IoT devices); and (3) auxiliary features that were collected by actively querying devices and were only used for evaluation. Embeddings were generated by converting the extracted features into key-value pairs and pseudo natural language and served as input to a deep fusion model. As the classifier is dependent on BC/MC packets, it is feasible mainly for Wi-Fi networks and hence requires that data be collected from each Wi-Fi access point which becomes a costly endeavor in corporate environments due to the presence of multiple access points. This approach also requires subject matter experts to identify the BC/MC protocols and features to be monitored, which is both expensive and error prone; moreover, more time and data are needed to identify IoT devices in this case.

\section{Proposed Methodology}\label{sec3}

\subsection{Approach}\label{subsec_3_1}
The approaches currently used to classify network communication are the statistical and behavioral feature-based approach and rule-based approach \citep{nguyen2008survey,zhang2014robust}. The limitation of the statistical and behavioral feature-based approach is that there is a need to identify the right features and preprocess them before inputting them in a machine learning model; thus this approach requires in-depth domain knowledge.  As the rule-based approach relies heavily on the port number of the network communication, it cannot be trusted.

In this research, we propose an approach based on representation learning for the identification of IoT devices in the network communication. We conducted five experiments to demonstrate the effectiveness of the proposed approach in different scenarios. The first is aimed at identifying whether the communication originates from IoT devices or non-IoT devices; in the second and third experiments, we focus respectively on identifying the communication of a particular IoT device within the IoT device network communication and within all of the network communication (including both IoT and non-IoT devices). In the fourth experiment, we identify multiple IoT devices simultaneously, along with the traffic of the non-IoT devices. The motive of the fifth experiment was to identify IoT devices that are not on the white-list (unknown devices) in the organizational network. The intuition behind this research is to leverage the short length and particular patterns formed when IoT devices communicate, which differ from that of smartphones and computers that use other protocols and have variable data lengths, in order to classify IoT devices in the network communication. The scope of our work is limited to IoT devices that use the TCP protocol for the connection. However, as our approach does not rely on any TCP protocol features, but rather uses the complete payload of the network communication, it is applicable (with minor modification) on IoT devices that use other protocols for connection, including Zigbee, CoAP, and Bluetooth.

\subsection{Data Preprocessing}\label{subsec_3_2}

As a part of the preprocessing step, we convert the network communication available in pcap (packet capture) format to grayscale images. The main focus in our approach is on the payloads of the TCP sessions, which are transferred between IoT devices, as shown in Figure \ref{fig:figure_2}.  The data preprocessing step is the same for all of the experiments (Table \ref{tab:tab2} contains details about the experiments).

The TCP session’s payload is converted to an image using the following steps \citep{kotak2019iot}:

\textbf{Step 1:} From the single pcap file, multiple pcap files are created based on the network sessions (i.e., a single network session is a group of packets that have the same source and destination IP address, source and destination port number, and protocol). After performing this step, each pcap file denotes a single network session. Tools similar to SplitCap \citep{netresec} can be used to perform this step. As the scope is restricted to the TCP protocol, UDP session data are ignored.

\textbf{Step 2:} After obtaining multiple pcap files (by performing step 1), the files are separated into different groups based on the source MAC address. Multiple folders are formed, where each folder has pcap files originating from a particular MAC address. Thus, the different folders contain the communication of the different IoT devices used in our experiments, based on the source MAC addresses.

\textbf{Step 3}: In this step, the header of each packet is removed, and therefore, we extract the TCP payload (in hexadecimal form) from every pcap file and save it in bin file format (binary file). Duplicate files and files with no data are ignored in this step.

\textbf{Step 4:} We fix the file size at 784 bytes so that each bin file will be of the same size. If the file size exceeds 784 bytes, we save just the first 784 bytes of data, and we append 0x00 bytes to the bin file if the file size is less than 784 bytes. 

\begin{figure}[h!]
   \centering
   \includegraphics[width=0.7\linewidth, keepaspectratio]{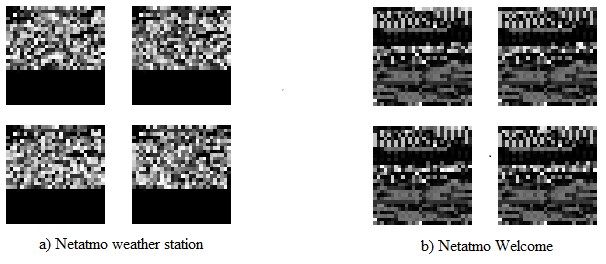}
   \caption{Visualization of the communication patterns of two different IoT devices}
   \label{fig:figure_1}
\end{figure}

\begin{figure*}[h!]
    \centering
    \includegraphics[width=0.7\linewidth, keepaspectratio]{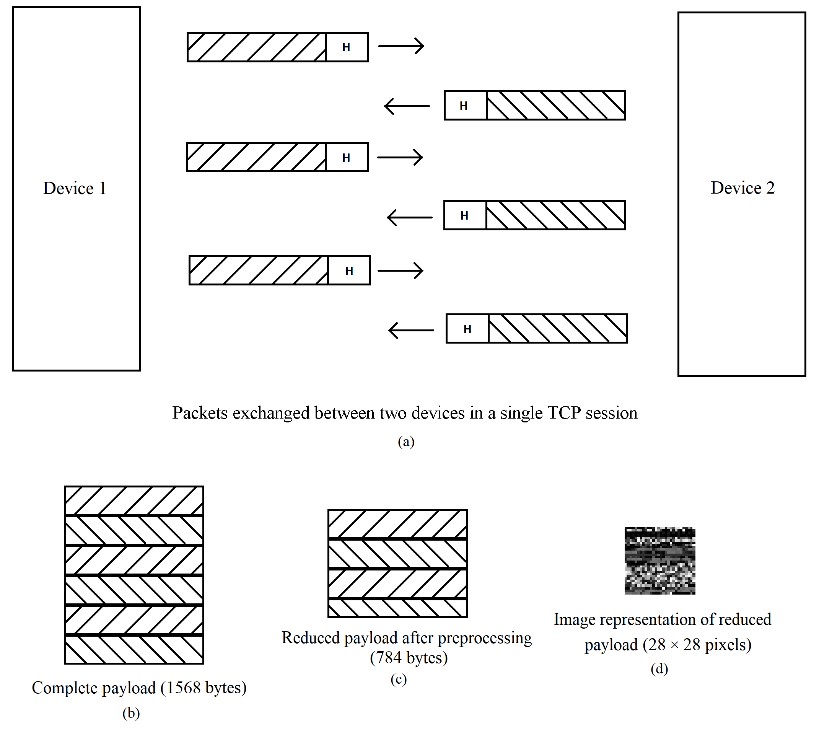}
    \caption{Converting a network communication payload into an image representation \citep{kotak2019iot}}
    \label{fig:figure_2}
\end{figure*}

\textbf{Step 5:} The fifth step is an optional step. As shown in Figure \ref{fig:figure_1}, in this step, 28 × 28 pixel grayscale images in which each pixel of an image consists of two hexadecimal digits (in total, 784 bytes consists of 1,568 hexadecimal digits), are generated from the bin files. To simplify the deep learning model’s training, IDX files are generated, similar to MNIST dataset files \citep{mnist}, from each class (IoT device).

Figure \ref{fig:figure_1} shows four random images created, based on the steps described above, for a Netatmo weather station IoT device and a Netatmo Welcome IoT device. As can be seen in the images, each IoT device has a discrete communication pattern. 

\begin{table*}[]
\centering
    \caption{Details about the dataset used in our experiments \citep{kotak2019iot}}
    \label{tab:tab1}
\resizebox{\linewidth}{!}{%
\begin{tabular}{@{}llrrrr@{}}
\toprule
\multicolumn{1}{c}{\textbf{Device Name}} & \multicolumn{1}{c}{\textbf{Device Type}} & \multicolumn{1}{p{0.15\textwidth}}{\textbf{Sessions in Training Set}} & \multicolumn{1}{p{0.15\textwidth}}{\textbf{Sessions in Validation Set}} & \multicolumn{1}{p{0.15\textwidth}}{\textbf{Sessions in Test Set}} & \multicolumn{1}{p{0.10\textwidth}}{\textbf{Total Sessions}} \\ \midrule

Samsung SmartCam                 & IoT                  & 7,313                                                 & 903                                                     & 813                                               & 9,029                                       \\
Withings Aura smart sleep sensor & IoT                  & 2,903                                                 & 358                                                     & 323                                               & 3,584                                       \\
Insteon camera                   & IoT                  & 3,285                                                 & 405                                                     & 365                                               & 4,055                                       \\
Amazon Echo                      & IoT                  & 2,759                                                 & 341                                                     & 307                                               & 3,407                                       \\
Netatmo weather station          & IoT                  & 1,894                                                 & 234                                                     & 210                                               & 2,338                                       \\
Netatmo Welcome                  & IoT                  & 2,177                                                 & 269                                                     & 242                                               & 2,688                                       \\
Pix-Star photo frame             & IoT                  & 906                                                   & 112                                                     & 100                                               & 1,118                                       \\
Belkin Wemo light switch         & IoT                  & 5,695                                                 & 703                                                     & 633                                               & 7,031                                       \\
Belkin Wemo motion sensor        & IoT                  & 31,199                                                & 3,852                                                   & 3,467                                             & 38,518                                      \\
Non-IoT devices                  & Non - IoT            & 20,035                                                & 2,474                                                   & 2,226                                             & 24,735                                      \\ \bottomrule
\end{tabular}}
\end{table*}

\subsection{Dataset and Environment}\label{subsec_3_3}

To validate the efficiency of our proposed approach, we trained a single-layer fully-connected neural network. The dataset considered in our experiments is part of the IoT Trace dataset \citep{sivanathan2018classifying}. The entire dataset has 218,657 TCP sessions (110,849 TCP sessions of IoT devices and 107,808 TCP sessions of non-IoT devices). Because our approach uses deep learning, we only considered devices with more than 1,000 TCP sessions for our experiments. The reduced dataset was further divided into three sets, i.e., training, validation, and test sets. The validation set was formed by randomly choosing 10\% of the data, and the test set was formed by randomly choosing 10\% from the remaining data. Table \ref{tab:tab1} contains details on the subset of the IoT Trace dataset used in our experiments (after performing the above mentioned preprocessing steps) \citep{kotak2019iot}. 

\begin{table*}[]
\centering
\caption{Summary of the experiments performed}
\label{tab:tab2}
\resizebox{\linewidth}{!}{%
\begin{tabular}{@{}p{5cm}clcp{8cm}@{}}
\toprule
\multicolumn{1}{p{3.7cm}}{\textbf{Experiment Number and Name}}                   & \multicolumn{1}{p{1.2cm}}{\textbf{Total Labels}} & \multicolumn{1}{p{2cm}}{\textbf{Network Traffic Considered}} & \multicolumn{1}{p{2cm}}{\textbf{Number of Classifiers as Outcome}} & \multicolumn{1}{c}{\textbf{Description}}                                                                                                                        \\ \midrule
1. Detection of IoT and non-IoT device traffic                            & 2                                         & IoT + non-IoT                                           & 1                                                             & All IoT devices are assigned the same label, and all non-IoT devices are assigned a single different label.                                                     \\
2. Detection of a specific IoT device in the IoT   device network traffic & 2                                         & Only IoT                                                & 9                                                             & All IoT devices except one (the excluded device) are assigned the same label, and the excluded device is assigned a different label in rotation.                \\
3. Detection of a specific IoT device in the network   traffic            & 2                                         & IoT + non-IoT                                           & 9                                                             & All network devices except one IoT device (the excluded device) are assigned the same label, and the excluded device is assigned a different label in rotation. \\
4. Detection of multiple IoT devices in the network   traffic             & 10                                        & IoT + non-IoT                                           & 1                                                             & Each IoT device is assigned a separate label, and all non-IoT devices are assigned a single different label.                                                    \\
5. Detection of unauthorized IoT devices in the   network traffic         & 8+1                                       & Only IoT                                                & 9                                                             & A single IoT device is excluded (only) from training, and the other eight IoT devices are assigned separate labels in rotation.                                 \\ \bottomrule
\end{tabular}}
\end{table*}

\subsection{Model Architecture}\label{subsec_3_4}

\begin{figure*}[h!]
    \centering
    \includegraphics[width=0.7\linewidth, keepaspectratio]{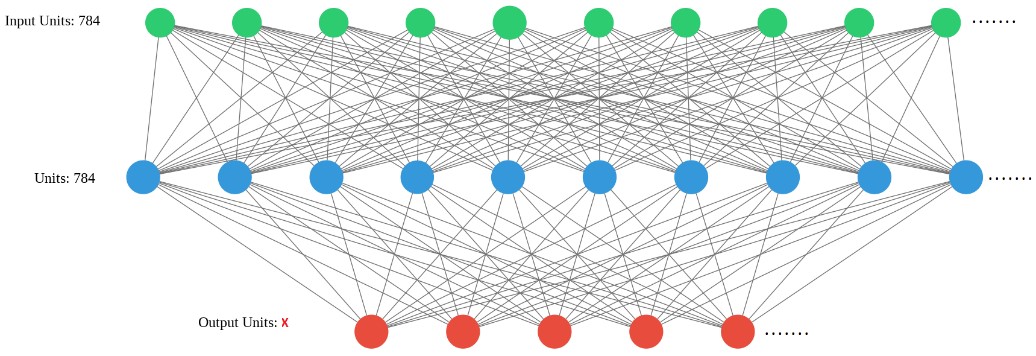}
    \caption{Neural network model architecture}
    \label{fig:figure3}
\end{figure*}

\begin{figure*}[h!]
    \centering
    \includegraphics[width=0.7\linewidth, keepaspectratio]{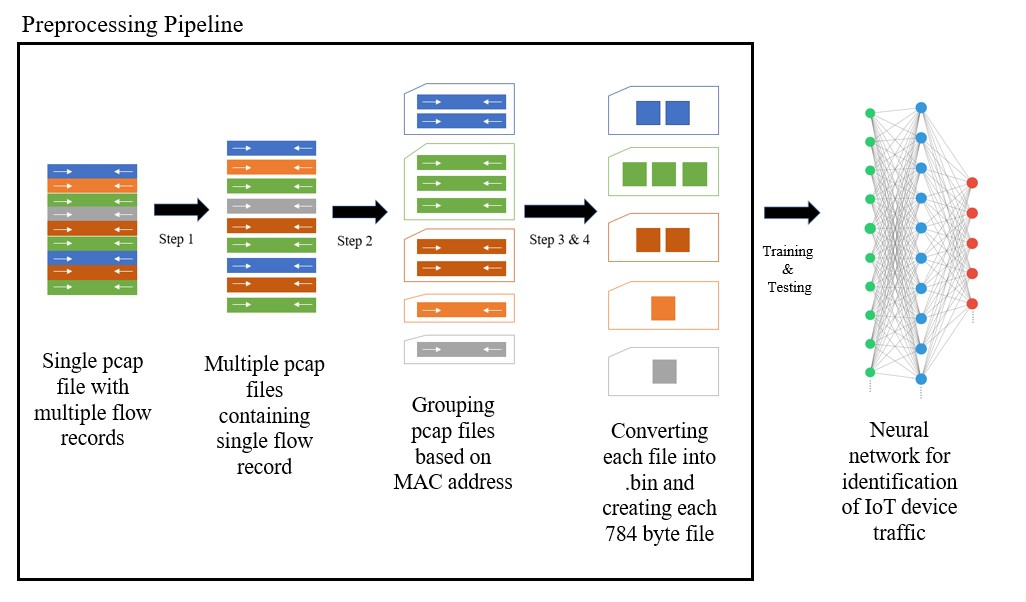}
    \caption{Entire dataflow pipeline}
    \label{fig:figure4}
\end{figure*}

Our proposed model has just one input layer and one output layer, making our approach simpler in terms of architecture than existing approaches for classifying IoT devices.  For all of the experiments, the input is a 28 × 28 pixel (i.e., 784 pixel value) grayscale image; if byte values are provided directly, the input layer should have 784 neurons. As we are classifying whether the communication originates from IoT or non-IoT devices in the first experiment, the output layer has one neuron. In the second experiment, we identify the communication of a particular IoT device within the IoT device communication, whereas, in the third experiment, we identify the communication of a particular IoT device within the network communication, including non-IoT device communication (i.e., IoT and non-IoT device communication). The output layer for the second and third experiments has one neuron. In the fourth experiment, in which we classify nine IoT devices along with one non-IoT device class, the output layer has 10 neurons. In our fifth experiment, in which we detect an unknown IoT device (which is not part of the white-list), we consider the communication of nine IoT devices and train nine models with similar architectures; one class of IoT device communication is kept out of the training set each time, and we consider the omitted IoT device communication as an unknown IoT device to demonstrate the usefulness of our approach. Therefore, the output layer of the fifth experiment has eight neurons. The neural network architecture of all of the experiments is presented in Figure \ref{fig:figure3}, where the X in the last layer represents the number of neurons. The five experiments are summarized in Table \ref{tab:tab2}. The model’s entire data-flow pipeline is presented in Figure \ref{fig:figure4}.

Weights were initialized for the input and output layers of the model using a \textit{normal} distribution. The activation function used in the input layer for all of the experiments was \textit{ReLU}; the activation function in the output layer for the first three experiments was \textit{sigmoid}, and for the last two experiments it was \textit{softmax}. \textit{Adam} was used as an optimization algorithm, while for the loss and the evaluation metric, \textit{categorical cross-entropy} and \textit{accuracy} were used respectively for the validation set \citep{team_init,team_act,team_opt,team_loss,team_met}. The results did not vary statistically significantly when additional intermediate hidden layers with different parameters were used.

\subsection{Evaluation Metrics}\label{subsec_3_5}
To validate our models on the test set, accuracy (A), precision (P), recall (R), and the F1 score (F1) were used as evaluation metrics. 

\section{Evaluation}\label{sec4}

\subsection*{Experiment 1: Detection of IoT and non-IoT devices}\label{subsec_4_1}

For this experiment, communication from all of the IoT devices was merged and assigned a single label. Hence, a binary classifier was trained, as there were only two labels - one for IoT device communication and another for non-IoT device communication.

\begin{figure*}[h!]
    \centering
    \includegraphics[width=0.8\linewidth, keepaspectratio]{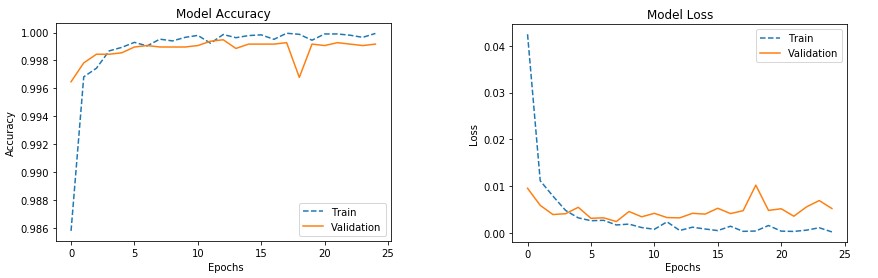}
    \caption{Model accuracy and loss on the validation set for the first experiment}
    \label{fig:figure5}
\end{figure*}

Figure \ref{fig:figure5} presents the model accuracy and loss obtained on the validation set after training a binary classifier for 25 epochs, with a batch size of 100. As shown in the graphs, the optimum validation accuracy (99.95\%) and validation loss were reached after 13 epochs. We obtained accuracy of 99.90\% on the test set after retraining the model (to avoid overfitting) for 13 epochs, which is higher than or equal to the accuracy of existing approaches. The performance evaluation results for the IoT devices and non-IoT devices are presented in Table \ref{tab:tab3}.
\begin{table}[]
\centering
\caption{Performance evaluation results for the first experiment}
\label{tab:tab3}
\resizebox{\columnwidth}{!}{%
\begin{tabular}{@{}p{2cm}p{1.5cm}rrrrr@{}}
\toprule
\textbf{Actual devices / Classified as} & \multicolumn{2}{p{1.5cm}}{\textbf{Non-IoT Devices}} & \textbf{IoT Devices} & \textbf{Precision} & Recall & F1 Score \\ \midrule
Non-IoT   Devices                         & \multicolumn{2}{r}{2,225}                      & 1                      & 0.997              & 1      & 0.998    \\
IoT   Devices                             & \multicolumn{2}{r}{7}                          & 6,453                  & 1                  & 0.999  & 0.999    \\
\multicolumn{2}{l}{}                                               & \multicolumn{2}{r}{Weighted Avg}             & 0.999              & 0.999  & 0.999    \\ \bottomrule
\end{tabular}}
\end{table}

\subsection*{Experiments 2 and 3: Detection of a specific IoT device }\label{subsec_4_2}

In the second experiment, non-IoT devices were not considered. Two labels were used - one label for the IoT device that we want to identify and another label for the other IoT devices. For each IoT device, Table \ref{tab:tab4} presents the minimum number of epochs required to obtain the optimum validation accuracy and validation loss for the classifiers and the test set accuracy obtained after retraining the classifiers using their respective minimum number of epochs. The performance evaluation results for each IoT device classifier in this experiment are presented in Appendix \ref{secA1}.

\begin{table}[]
\caption{Minimum number of epochs and test set accuracy obtained for each classifier in the second experiment}
\label{tab:tab4}
\resizebox{\columnwidth}{!}{%
\begin{tabular}{@{}p{4.5cm}rr@{}}
\toprule
\multicolumn{1}{c}{\textbf{IoT Device Classifier}} & \multicolumn{1}{p{1.8cm}}{\textbf{Minimum Number of Epochs}} & \multicolumn{1}{p{1.75cm}}{\textbf{Test Set Accuracy (\%)}} \\ \midrule
Samsung SmartCam                                   & 3                                                     & 100                                                 \\
Withings Aura smart sleep sensor                   & 3                                                     & 99.9                                                \\
Insteon camera                                     & 3                                                     & 99.9                                                \\
Amazon Echo                                        & 5                                                     & 99.9                                                \\
Netatmo weather station                            & 1                                                     & 100                                                 \\
Netatmo Welcome                                    & 1                                                     & 99.9                                                \\
Pix-Star photo frame                               & 1                                                     & 100                                                 \\
Belkin Wemo light switch                           & 10                                                    & 99.9                                                \\
Belkin Wemo motion sensor                          & 3                                                     & 99.8                                                \\ \bottomrule
\end{tabular}}
\end{table}
 
In the third experiment, all of the devices were considered, including non-IoT devices. There were two labels - one label for the IoT device that we want to identify and another label for the other devices (including both IoT and non-IoT devices). As the second and third experiments had two labels, binary classifiers were trained.  For each IoT device, Table \ref{tab:tab5} presents the minimum number of epochs required to obtain optimum validation accuracy and validation loss for the classifiers and the test set accuracy obtained after retraining the classifiers using their respective minimum number of epochs. The performance evaluation results for each IoT device classifier in this experiment are presented in Appendix \ref{secA2}.

\begin{table}[]
\caption{Minimum number of epochs and test set accuracy obtained for each classifier in the third experiment}
\label{tab:tab5}
\resizebox{\columnwidth}{!}{%
\begin{tabular}{@{}p{4.5cm}rr@{}}
\toprule
\multicolumn{1}{c}{\textbf{IoT Device Classifier}} & \multicolumn{1}{p{1.8cm}}{\textbf{Minimum Number of Epochs}} & \multicolumn{1}{p{1.75cm}}{\textbf{Test Set Accuracy (\%)}} \\ \midrule
Samsung SmartCam                                   & 5                                                     & 100                                                 \\
Withings Aura smart sleep sensor                   & 7                                                     & 99.9                                                \\
Insteon camera                                     & 1                                                     & 99.9                                                \\
Amazon Echo                                        & 5                                                     & 99.9                                                \\
Netatmo weather station                            & 1                                                     & 100                                                 \\
Netatmo Welcome                                    & 4                                                     & 100                                                \\
Pix-Star photo frame                               & 1                                                     & 100                                                 \\
Belkin Wemo light switch                           & 4                                                    & 99.9                                                \\
Belkin Wemo motion sensor                          & 19                                                     & 99.8                                                \\ \bottomrule
\end{tabular}}
\end{table}

\subsection*{Experiment 4: Detection of multiple IoT devices}\label{subsec_4_3}
For this experiment, all of the devices were considered (nine IoT devices and the non-IoT devices). Nine labels were assigned to the nine IoT devices, and all non-IoT devices were assigned a single different label. As there were multiple labels, a multiclass classifier was trained \citep{kotak2019iot}.
\begin{figure*}[h!]
    \centering
    \includegraphics[width=0.8\linewidth, keepaspectratio]{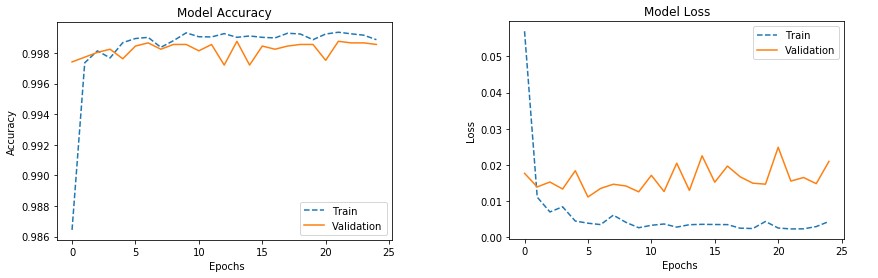}
    \caption{Model accuracy and loss on the validation set for the fourth experiment \citep{kotak2019iot}}
    \label{fig:figure6}
\end{figure*}
Figure \ref{fig:figure6} presents the model accuracy and loss obtained on the validation set after training a multiclass classifier for 25 epochs, with a batch size of 100. As shown in the graphs, the optimum validation accuracy (99.87\%) and validation loss were obtained after seven epochs. We obtained accuracy of 99.86\% on the test set after retraining the model (to avoid overfitting) for seven epochs, which is higher than or equal to the accuracy of existing approaches. The performance evaluation results for each IoT device and the non-IoT devices are presented in Table \ref{tab:tab6}.

The k-fold cross-validation evaluation results for first four experiments are presented in Appendix \ref{secA4}.

\begin{table*}[]
\caption{Performance evaluation results for the fourth experiment \citep{kotak2019iot}}
\label{tab:tab6}
\resizebox{\linewidth}{!}{%
\begin{tabular}{@{}lrrrrrrrrrrrrr@{}}
\toprule
\multicolumn{1}{c}{\textbf{Actual device / Classified as}} & \multicolumn{1}{c}{\textbf{0}} & \multicolumn{1}{c}{\textbf{1}} & \multicolumn{1}{c}{\textbf{2}} & \multicolumn{1}{c}{\textbf{3}} & \multicolumn{1}{c}{\textbf{4}} & \multicolumn{1}{c}{\textbf{5}} & \multicolumn{1}{c}{\textbf{6}} & \multicolumn{1}{c}{\textbf{7}} & \multicolumn{1}{c}{\textbf{8}} & \multicolumn{1}{c}{\textbf{9}} & \multicolumn{1}{c}{\textbf{P}} & \multicolumn{1}{c}{\textbf{R}} & \multicolumn{1}{c}{\textbf{F1}} \\ \midrule
0- Non-IoT devices                                         & 2,226                          & 0                              & 0                              & 0                              & 0                              & 0                              & 0                              & 0                              & 0                              & 0                              & 0.998                          & 1                              & 0.999                           \\
1-   Samsung SmartCam                                      & 1                              & 812                            & 0                              & 0                              & 0                              & 0                              & 0                              & 0                              & 0                              & 0                              & 1                              & 0.999                          & 0.999                           \\
2-   Withings Aura smart sleep sensor                      & 2                              & 0                              & 321                            & 0                              & 0                              & 0                              & 0                              & 0                              & 0                              & 0                              & 1                              & 0.994                          & 0.997                           \\
3-   Insteon camera                                        & 0                              & 0                              & 0                              & 365                            & 0                              & 0                              & 0                              & 0                              & 0                              & 0                              & 1                              & 1                              & 1                               \\
4-   Amazon Echo                                           & 0                              & 0                              & 0                              & 0                              & 306                            & 1                              & 0                              & 0                              & 0                              & 0                              & 1                              & 0.997                          & 0.998                           \\
5-   Netatmo weather station                               & 0                              & 0                              & 0                              & 0                              & 0                              & 210                            & 0                              & 0                              & 0                              & 0                              & 0.995                          & 1                              & 0.998                           \\
6- Netatmo   Welcome                                       & 1                              & 0                              & 0                              & 0                              & 0                              & 0                              & 241                            & 0                              & 0                              & 0                              & 1                              & 0.996                          & 0.998                           \\
7- Pix-Star   photo frame                                  & 0                              & 0                              & 0                              & 0                              & 0                              & 0                              & 0                              & 100                            & 0                              & 0                              & 1                              & 1                              & 1                               \\
8-   Belkin Wemo light switch                              & 0                              & 0                              & 0                              & 0                              & 0                              & 0                              & 0                              & 0                              & 628                            & 5                              & 0.998                          & 0.992                          & 0.995                           \\
9-   Belkin Wemo motion sensor                             & 1                              & 0                              & 0                              & 0                              & 0                              & 0                              & 0                              & 0                              & 1                              & 3,465                          & 0.999                          & 0.999                          & 0.999                           \\
                                                           &                                &                                &                                &                                &                                &                                & \multicolumn{4}{r}{Weighted   Avg}                                                                                                & 0.999                          & 0.999                          & 0.999                           \\ \bottomrule
\end{tabular}}
\end{table*}

\subsection*{Experiment 5: Detection of unauthorized IoT devices}\label{subsec_4_3}
For this experiment, the dataset included the communication of nine IoT devices. Nine different multiclass classifiers were trained; each time, communication from eight IoT devices was used to train the classifier, excluding the communication of one IoT device that was labeled as an unknown IoT device. The minimum number of epochs required by each classifier to obtain the maximum accuracy on the validation set was determined after training each classifier for 30 epochs \citep{kotak2019iot}.

Each classifier was retrained with its respective minimum number of epochs to avoiding overfitting. When any single instance is provided to a classifier as input, the classifier outputs a single array of posterior probabilities with a length of eight. Each probability value indicates the possibility of the given instance to have originated from one of the eight IoT devices. The threshold value was obtained for the classification of an instance in such a way that, given the vector of probabilities, if any single probability that surpasses the threshold value exists, the instance is classified as originating from one of the eight IoT devices, based on the index of the probability in the output vector; otherwise, that instance is classified as unknown. The threshold values were obtained utilizing the validation set on the trained classifiers, which maximizes accuracy (A). Table \ref{tab:tab7} contains the threshold values for each classifier. A greater threshold value means that a classifier can differentiate unknown devices with greater confidence. The performance evaluation results for each classifier are presented in Appendix \ref{secA3}.

\begin{table}[]
\caption{Minimum number of epochs, thresholds, and test set accuracy obtained for each classifier in the fifth experiment \citep{kotak2019iot}}
\label{tab:tab7}
\resizebox{\columnwidth}{!}{%
\begin{tabular}{@{}lrrr@{}}
\toprule
\multicolumn{1}{c}{\textbf{Unknown   Device}} & \multicolumn{1}{p{1.5cm}}{\textbf{Minimum   Number of Epochs}} & \multicolumn{1}{p{1.5cm}}{\textbf{Thresholds}} & \multicolumn{1}{p{1.5cm}}{\textbf{Test   Set Accuracy (\%)}} \\ \midrule
Samsung   SmartCam                            & 9                                                       & 0.97                                    & 98.9                                                  \\
Withings   Aura smart sleep sensor            & 27                                                      & 0.99                                    & 97.9                                                  \\
Insteon   camera                              & 5                                                       & 0.77                                    & 99.3                                                  \\
Amazon   Echo                                 & 18                                                      & 0.99                                    & 98.3                                                  \\
Netatmo   weather station                     & 8                                                       & 0.92                                    & 98.8                                                  \\
Netatmo   Welcome                             & 6                                                       & 0.80                                    & 99.8                                                  \\
Pix-Star   photo frame                        & 3                                                       & 0.76                                    & 99.8                                                  \\
Belkin   Wemo light switch                    & 3                                                       & 0.87                                    & 99.8                                                  \\
Belkin   Wemo motion sensor                   & 3                                                       & 0.90                                    & 99.0                                                  \\ \bottomrule
\end{tabular}}
\end{table}

\section{Discussion}\label{sec5}
We achieved 99\% overall average accuracy in detecting IoT devices based on the network payload in five different scenarios that could take place in organizations. Our comprehensive evaluation demonstrated the effectiveness of the proposed approach and its ability to identify both unauthorized IoT devices and white-listed devices and serve as a tool for enforcing organizations’ security policies. Unlike other approaches, the proposed approach can also correctly identify different IoT devices from the same vendors \citep{miettinen2017iot}. The use of a NAT, as is done in many networks, can cause other approaches to fail \citep{meidan2017detection}, however our approach utilizes the initial bytes of a payload which are not affected by a NAT and therefore works in NAT-enabled environments. In addition, our generic approach, which can be applied in many scenarios, has the advantages of simpler preprocessing steps and model architecture, and greater effectiveness than existing approaches for IoT identification \citep{meidan2017profiliot,sivanathan2017characterizing,aksoy2019automated,miettinen2017iot,sivanathan2018classifying,yu2020you}. 

\section{Conclusion}\label{sec6}
In this research, we proposed an approach that utilizes deep learning to identify both known and unauthorized IoT devices within the network communication. We demonstrated the efficiency of our approach in different scenarios, attaining over 99\% overall average accuracy. We also explained the key advantages of our approach, including its simple architecture (unlike existing approaches) and the fact that it needs no feature engineering (eliminating the associated overhead). As our approach concentrates on the network communication payload instead of the packet header fields, it is applicable for any IoT device, regardless of the protocol utilized for communication and therefore is more generic. For the same reason, it is applicable for the IoT devices behind a NAT-enabled router. As part of future research, we plan to apply our approach in other scenarios, including scenarios involving the use of various network protocols that are not based on the TCP/IP network stack for network communication.

\backmatter

\bmhead{Acknowledgments}

This project was partially funded by the European Union’s Horizon 2020 research and innovation programme under grant agreement No. 830927.

\section*{Declarations}

\textbf{Conflict of interest:} The authors declare that they have no conflicts of interest.

\onecolumn
\begin{appendices}

\section{Performance evaluation results for the detection of a specific IoT device in the IoT device network communication}\label{secA1}
\begin{figure*}[h!]
    \centering
    \includegraphics[width=1\linewidth, keepaspectratio]{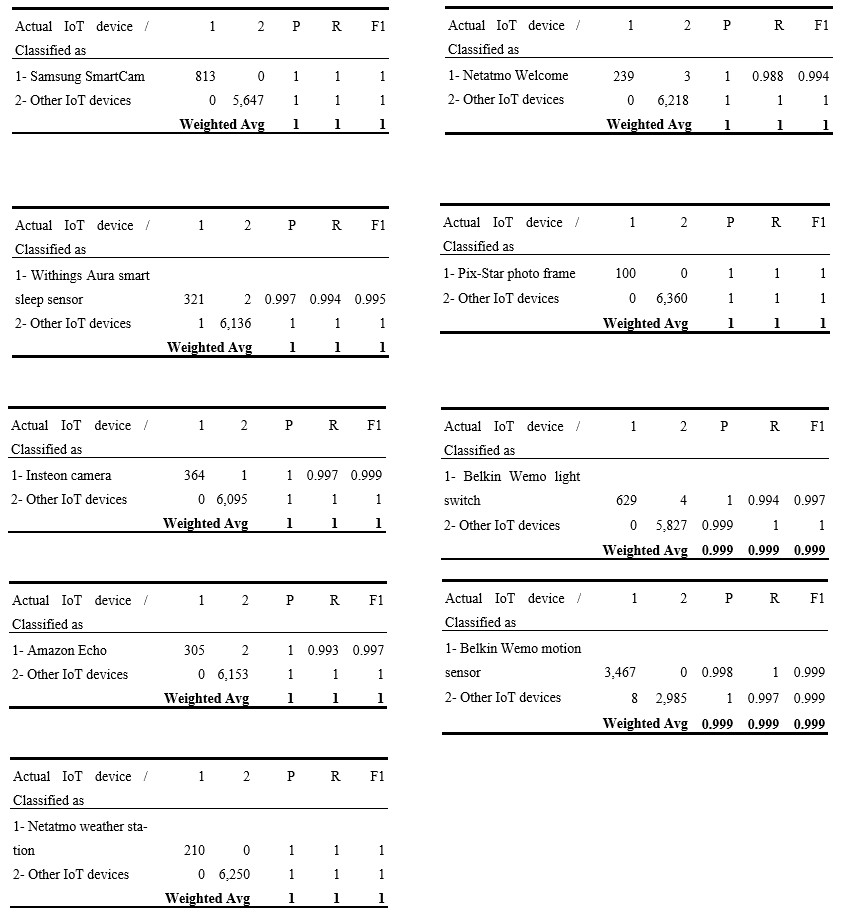}
    \label{fig:a1}
\end{figure*}

\section{Performance evaluation results for the detection of a specific IoT device in the network communication}\label{secA2}
\begin{figure*}[h!]
    \centering
    \includegraphics[width=1\linewidth, keepaspectratio]{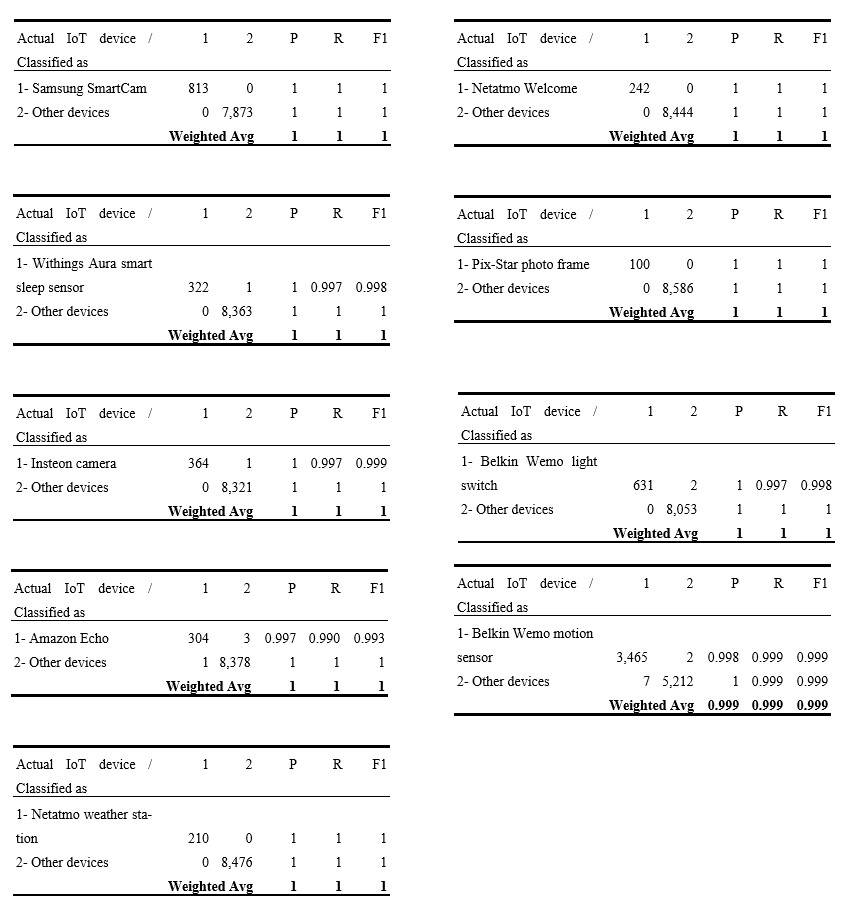}
    \label{fig:a1}
\end{figure*}

\section{Performance evaluation results for the detection of unauthorized IoT devices}\label{secA3}
\begin{figure*}[h!]
    \centering
    \includegraphics[width=1\linewidth, keepaspectratio]{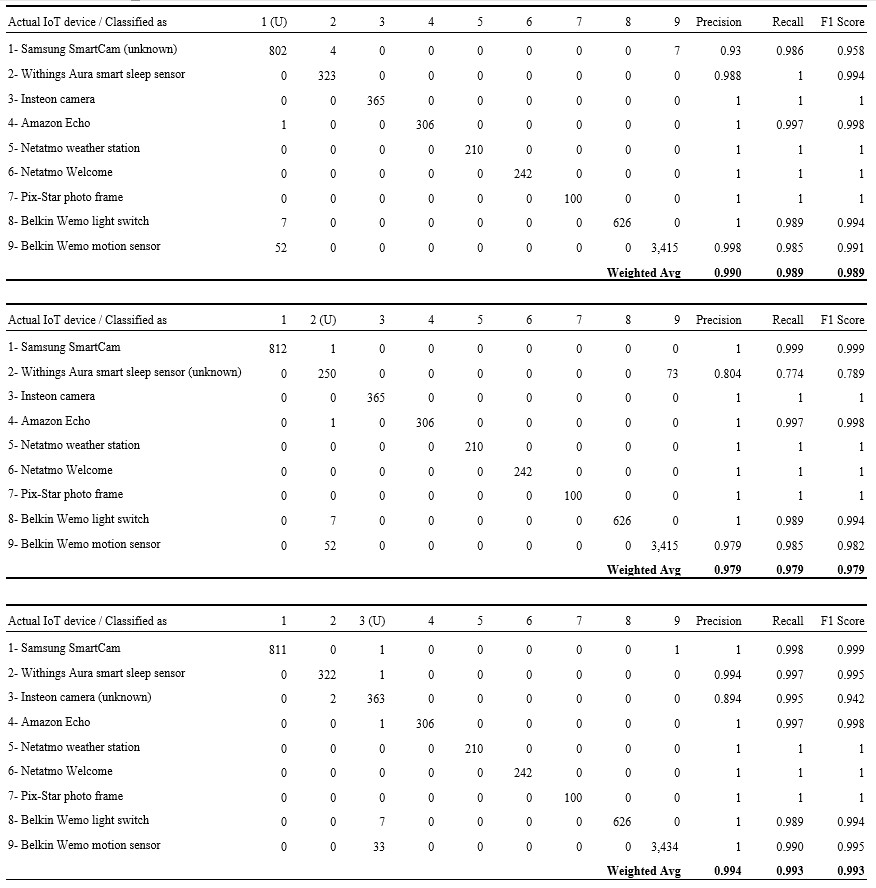}
    \label{fig:a1}
\end{figure*}
\begin{figure*}[h!]
    \centering
    \includegraphics[width=1\linewidth, keepaspectratio]{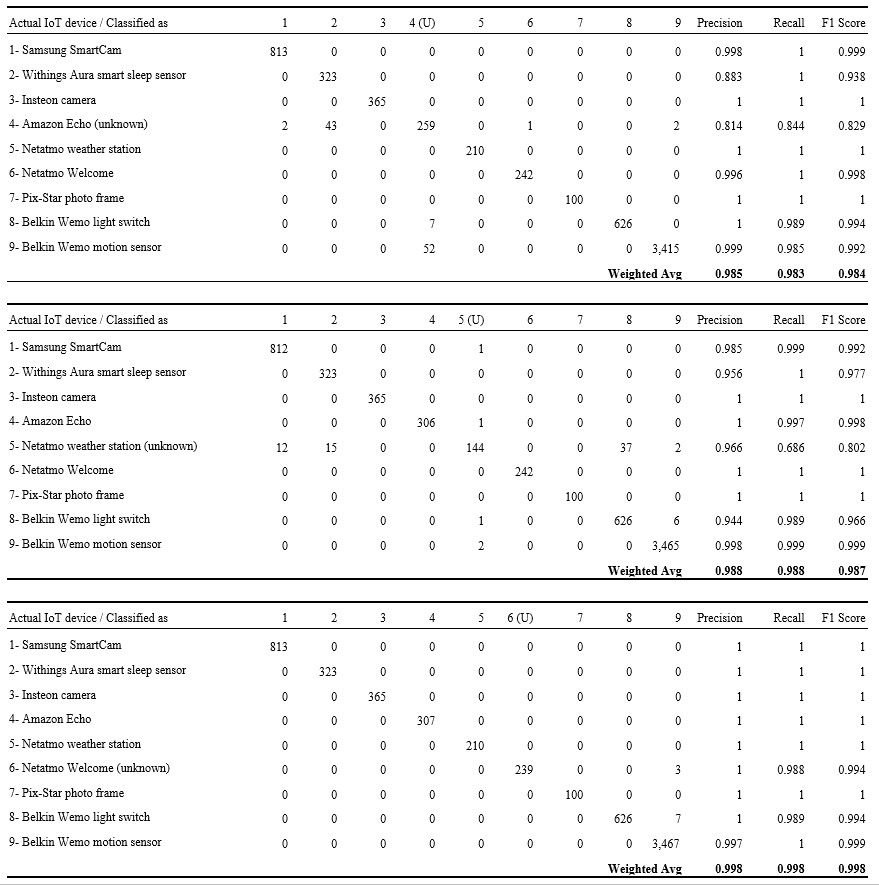}
    \label{fig:a1}
\end{figure*}
\clearpage
\begin{figure*}[h!]
    \centering
    \includegraphics[width=1\linewidth, keepaspectratio]{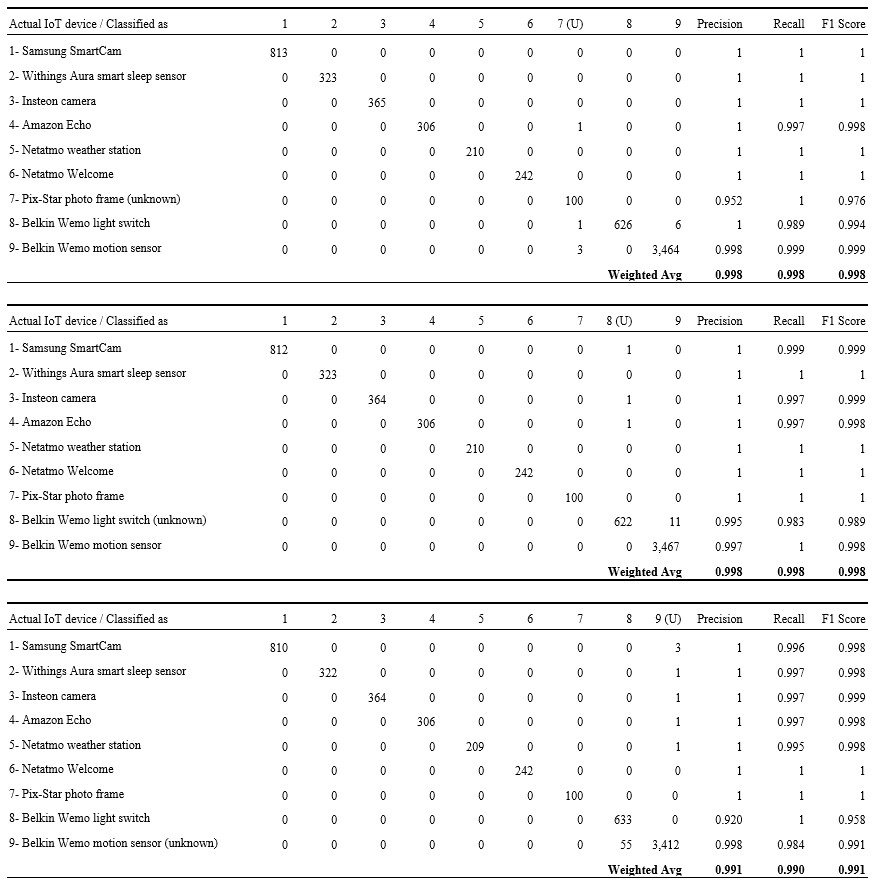}
    \label{fig:a1}
\end{figure*}
\clearpage

\section{K-fold cross-validation results of different classifiers for the first four experiments}\label{secA4}
\begin{figure*}[h!]
    \centering
    \includegraphics[width=0.8\linewidth, keepaspectratio]{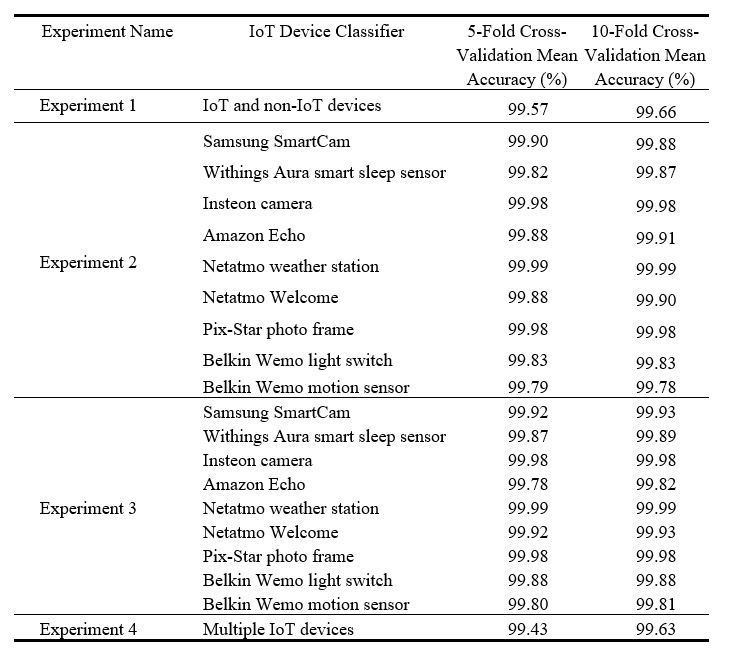}
    \label{fig:a1}
\end{figure*}

* the k-fold cross-validation results for the fifth experiment are not listed, as it requires deriving a threshold from the validation set which is not feasible in k-fold cross-validation.




\end{appendices}
\twocolumn
\bibliographystyle{plainnat}
\bibliography{sn-bibliography}


\end{document}